\documentclass[10pt, a4paper]{article}
\usepackage[english]{babel}
\usepackage{amssymb,amsxtra} 
\usepackage[adobe-utopia]{mathdesign}

\newcommand{\reell}{\kern+.23em\sf{1}\kern-.61em\sf{1}\kern+.76em\kern-.25em}

\usepackage{graphicx,color}
\usepackage{makeidx}
\usepackage{url}
\usepackage{xcolor}
\DeclareMathVersion{teste}

\usepackage{tcolorbox}

\title{\textbf{Mass in the Light of Special Relativity}}

\author{
    Francisco Caruso \\
    \vspace*{-0.1cm}
    \small Centro Brasileiro de Pesquisas F\'{\i}sicas -- CBPF \\
    \vspace*{-0.1cm}
    \small Coordena\c{c}\~{a}o de F\'{\i}sica de Altas Energias \\
    \vspace*{-0.1cm}
    \small 22290-180, Rio de Janeiro, RJ, Brasil
    \and
    Vitor Oguri  \\ 
    \vspace*{-0.1cm}
    \small Universidade do Estado do Rio de Janeiro -- UERJ \\
    \vspace*{-0.1cm}
    \small Departamento de F\'{\i}sica Nuclear e Altas Energias -- DFNAE \\
    \vspace*{-0.1cm}
    \small 20550-900, Rio de Janeiro, RJ, Brasil
}

\date{}

\begin{document}

\mathversion{teste}

\markboth{ }{ }
\baselineskip 13.1pt

\parskip=4pt

\setcounter{equation}{0}





\maketitle

\vspace*{-0.6cm}
\begin{abstract}
The text analyzes the concept of mass within Special Relativity, initially deconstructing the interpretation of Einstein's equation ($\varepsilon_\circ = m c^2$) as a principle of equivalence. Based on the adoption of the conservation laws of momentum and energy, the non-additivity of mass in composite systems is established. While in atomic and nuclear systems the total mass is less than the sum of its constituents, in the subnuclear domain the opposite occurs: the positive potential energy of confinement of quarks causes the mass of a hadron to be greater than the sum of the masses of its quarks; in the case of nucleons, it is much greater. It is concluded, then, that the mass of baryonic matter in the observable Universe comes, for the most part, from the energy associated with strong interactions.


\end{abstract}

\vspace*{0.2cm}

\hrule
\vspace*{0.2cm}

\begin{flushright}
\begin{minipage}{7.5cm}
\baselineskip=10pt {\small
\textit{A generalization made not for the vain pleasure of generalizing but in order to solve previously existing problems is always a fruitful generalization.}
\smallskip

\hfill Henri Lebesgue}
\end{minipage}
\end{flushright}

\vspace*{-0.5cm}
\section{\textit{Momentum} and energy of a free particle in different inertial frames}

Modern physics theories have imposed radical changes on the view of  ``physical reality'' -- whether regarding not only the concepts of space and time, but also the hypothesis of Laplacian determinism; on the other hand, continuity is maintained in the formal structures of the theories of Special Relativity and Quantum Mechanics.

The formal similarity between modern and classical theories stems from analogies between certain phenomena of distinct natures, such as electromagnetic radiation and particle motion. These analogies led to the expansion and generalization of concepts originating in Classical Mechanics and Electromagnetism, such as particle, mass, electric charge, energy, and momentum.

According to Classical Mechanics, the energy ($\varepsilon$) and momentum ($\vec p$) of a free particle of mass $m$, relative to a given inertial frame $S$, are expressed and related as follows:
$$(\mbox{\small classical free particle}) \quad
\left\{
\begin{array}{l}
\displaystyle \varepsilon = \frac{1}{2} \, m\, v^2  \\
\ \\
\displaystyle \vec p =  m\, \vec v
\end{array}
\right. \qquad \Longrightarrow \qquad \varepsilon   = \frac{p^2}{2m} = \frac{1}{2}\, p \, v
$$
where $\vec v$ is the particle's velocity.

While Einstein [1] generalizes the concept of energy, defining the energy of a free particle as
$$\varepsilon =  \gamma (v)\, m\, c^2 \qquad \qquad (\mbox{\small Einstein -- free particle})$$
where $\gamma (v) = \big( 1 - v^2/c^2\big)^{-1/2} $ is the so-called Lorentz factor, and $c \simeq 3.0 \times 10^{8}$ m/s is the speed of light in a vacuum, it fell to Planck [2] to properly establish the equation of motion for an electrically charged particle in an electromagnetic field, and
to generalize the concept of particle momentum by identifying it as
$$ \displaystyle \vec p = \gamma(v)\,  m\, \vec v \qquad \qquad (\mbox{\small Planck})$$
\indent
The relativistic expression for the energy of a free particle implies that its value is never zero. In a reference frame in which the free particle is at rest, the energy denoted by
\begin{equation}\label{eq_einstein2}
\varepsilon_\circ = m c^2
\end{equation}
is termed by Einstein the particle's \textit{rest energy}.

According to the definitions of Einstein and Planck,
the energy, the rest energy, the momentum, and the velocity of the free particle
are related as follows:\footnote{\, It is from equation (\ref{S_energia}) that the mass ($m$) of an unstable particle with electric charge $e$e and a very short lifetime is determined. Momentum ($p$) is calculated by measuring the radius of curvature ($r$) of the particle's trajectory in a given magnetic field ($B$) using the relation $p = e r B$, and energy ($\varepsilon$) is measured independently in a calorimeter, such that mass ($m$) is determined by
$$ \displaystyle m = \sqrt{\varepsilon^2 - (pc)^2}/c^2$$  }
$$ 
\hspace*{2.7cm}  (\mbox{\small relativistic free particle}) \quad
\left\{
\hspace*{-3.5cm}
\begin{minipage}{12.7cm}
\vspace*{-0.3cm}
\begin{eqnarray}
 \qquad  \varepsilon^2  \hspace*{-0.2cm} & =&  \hspace*{-0.2cm}\displaystyle  (p c)^2 \ + \ \big( m c^2 \big)^2 =  (p c)^2 \ + \ (\varepsilon_\circ )^2 \label{S_energia} \\
  \ \nonumber \\
  \qquad \vec  p \hspace*{-0.2cm} &=  & \hspace*{-0.2cm}\displaystyle  \frac{\varepsilon}{c^2} \, \vec v  \label{pv_relat}
\end{eqnarray}
\end{minipage}
\right.
$$
%
\indent
Unlike the classical expression, the energy of a massive free particle -- in addition to the kinetic component that depends on velocity (and, consequently, on the reference frame) -- also possesses an intrinsic component associated with mass: rest energy.

\begin{figure}[hbtp]
\centerline{\includegraphics[width=5.5cm]{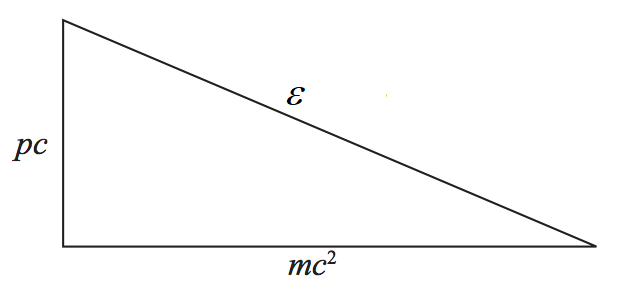}}
\vspace*{-0.3cm}
\caption{\small Relationship between the energy, momentum, and rest energy of a free particle.}
\label{tri_energia}
\end{figure}

In addition to massive particles, the expressions (\ref{S_energia}) and (\ref{pv_relat}) are also valid for massless particles, such as the photon. In these cases, the energy ($\varepsilon$) and the momentum ($p$) are related as
\begin{equation}\label{energia_foton}
\displaystyle \varepsilon = p\, c
\end{equation}
\indent As a mnemonic aid, the relationship between the energy, momentum, and rest energy of a massive free particle can be visualized as a composition that obeys the Pythagorean theorem (Figure~\ref{tri_energia}).
%

How, however, can the energy and momentum of a free particle in different inertial reference frames be related?

In another inertial frame $S'$ moving at a constant velocity $V$ -- also in the same direction as the particle's motion relative to $S$ -- the particle's energy ($\varepsilon'$) and momentum ($\vec p^{\,\prime}$) are related by
\begin{equation}\label{Sp_energia}
\displaystyle \varepsilon'^2 = (p' c)^2 \ + \ \big( m c^2 \big)^2 =  (p' c)^2 \ + \ (\varepsilon_\circ )^2
\end{equation}


Since the mass -- and therefore the rest energy -- of a particle is an attribute independent of the reference frame, in accordance with equations (\ref{S_energia}) and (\ref{Sp_energia}), the relations between the energies and the momenta satisfy the equality
\begin{equation} \label{mc2}
\displaystyle \varepsilon'^2 -  (p' c)^2  = \displaystyle \varepsilon^2 - (p c)^2
\end{equation}

By analogy with Lorentz transformations -- the linear transformations between the temporal and spatial coordinates, $(t, x)$ and $(t', x')$, of a particle in two distinct inertial reference frames ($S$ and $S'$) such that
$$ (c t')^2 - {x^\prime}^2 = (c t)^2 - x^2 $$
-- one can consider that energy and momentum in the reference frames $S$ and $S'$ are also related by linear equations, i.e.,
\begin{equation} \label{pe_trans}
\left\{
\begin{array}{l}
\displaystyle \varepsilon' = \alpha\, \varepsilon + \beta\, p  \\
\ \\
\displaystyle p' =  \delta\, \varepsilon + \gamma\, p
\end{array}
\right.
\end{equation}
where $\alpha$, $\beta$, $\delta$, and $\gamma$ are constants.

If $S'$ moves at the same velocity as the particle ($V = v$), according to equations~(\ref{pv_relat}) and (\ref{pe_trans}), it follows that
$$ \displaystyle p'=0 \qquad \Longrightarrow \qquad p = - \delta \frac{\varepsilon}{\gamma} = \varepsilon \frac{V}{c^2}
\qquad \Longrightarrow \qquad  \delta = - \gamma \frac{V}{c^2} $$
\indent Thus, equations~(\ref{pe_trans}) can be rewritten as
\begin{equation} \label{pe_trans_2}
\left\{
\begin{array}{l}
\displaystyle \varepsilon' = \alpha \left( \varepsilon + \frac{\beta}{\alpha} \, p \right)  \\
\ \\
\displaystyle p' =  \gamma \left( p - \frac{V}{c^2}\varepsilon \right)
\end{array}
\right.
\end{equation}
and the equality expressed by the equation~(\ref{mc2}),
\begin{eqnarray*}
\varepsilon'^2 - p'^2 c^2 & =& \alpha^2 \left( \varepsilon^2 + 2 \frac{\beta}{\alpha} \varepsilon \, p + \frac{\beta^2}{\alpha^2} p^2 \right) - \gamma^2 c^2 \left( p^2 - 2 \frac{V}{c^2}\, \varepsilon \, p + \frac{V^2}{c^4} \varepsilon^2 \right)\\
&=& \varepsilon^2 \Big(\underbrace{\alpha^2 - \gamma^2 \frac{V^2}{c^2}}_{1} \Big) + 2 \, \big(\underbrace{ \alpha \beta + \gamma^2 V }_{0} \big) \varepsilon \, p - p^2 c^2 \Big( \underbrace{\gamma^2 - \frac{\beta^2}{c^2}}_{1}\Big)
\end{eqnarray*}
is satisfied only if
$$
\left\{
\begin{array}{l}
\displaystyle \gamma^2 - 1 = \beta^2/c^2 \qquad \Longrightarrow \qquad \beta^2 = c^2( \gamma^2 -1) \\
\ \\
\displaystyle \alpha \beta = - \gamma^2 V \qquad \Longrightarrow \qquad \alpha^2 = \gamma^4 \frac{V^2}{\beta^2} \\
\ \\
\displaystyle \alpha^2 = 1 + \gamma^2 \frac{V^2}{c^2} = \gamma^4 \frac{V^2}{c^2} \frac{1}{\gamma^2 -1}
\end{array}
\right. \qquad \Longrightarrow \qquad
\left\{
\begin{array}{l}
\displaystyle \gamma^2 = \frac{1}{1 - V^2/c^2} \\
\ \\
\displaystyle \alpha^2 = 1 + \frac{V^2/c^2}{1 - V^2/c^2} = \gamma^2\\
\ \\
\displaystyle \beta^2 = \gamma^2 V^2
\end{array}
\right.
$$
\indent For $ V=0$, the equations (\ref{pe_trans}) imply
$$ \left\{ \begin{array}{l}
\beta|_{V=0} = \delta|_{ V=0}=0 \\
\ \\
\alpha|_{V=0} = \gamma|_{V=0} =1
\end{array} \right.
\quad \Longrightarrow \quad \alpha = \gamma = \displaystyle \frac{1}{\sqrt{1 - V^2/c^2}}
\quad \Longrightarrow \quad \beta = - \gamma V $$

Therefore, the relationships between the particle's energy and momentum in the two reference frames ($S$ and $S'$) can be expressed as\footnote{\, In the non-relativistic regime, according to Galilean velocity addition,
the particle's velocities ($v$ and $v'$) in the two reference frames ($S$ and $S'$) are related by $v' = v - V$, and therefore
the relationships between energy and momentum are expressed as:
$$\left\{
\begin{array}{l}
\displaystyle \varepsilon' =  \varepsilon -   p\, V + \frac{1}{2} m V^2\\
\ \\
\displaystyle p' =   p -  m V
\end{array}
\right.
$$}
\begin{equation} \label{pe_trans_3}
\left\{
\begin{array}{l}
\displaystyle \varepsilon' = \gamma(V)\, \big(  \varepsilon -   p\, V \big)  \\
\ \\
\displaystyle p' =  \gamma(V)\, \big( p -  \varepsilon\, V/c^2 \big) =  \gamma(V)\, \big[ p -  \gamma(v)  m  V \big]
\end{array}
\right.
\end{equation}
where $\gamma (V) = \big( 1 - V^2/c^2\big)^{-1/2} $ is the Lorentz factor.

If the particle's displacement is not parallel to the velocity of $S'$, according to Cartesian coordinate systems in $S$ and $S'$ with parallel axes, such that $S'$ moves in the positive direction of the $x$ axis, the components ($p_y$ and $p_z$) of momentum perpendicular to the velocity $V$ of $S'$ remain constant, while the component in the direction of $V$ ($p_x$) is modified, according to equation ~(\ref{pe_trans_3}). In this case, the transformation equations are given by
\begin{equation} \label{pe_trans_final}
\left\{
\begin{array}{l}
\displaystyle \varepsilon' = \gamma(V)\, \big( \varepsilon - p_x\, V \big) \\
\ \\
\displaystyle p_x^\prime = \gamma(V)\, \big( p_x - \varepsilon\, V/c^2 \big) \qquad \quad p_y^\prime = p_y
\qquad \quad p_z^\prime = p_z
\end{array}
\right.
\end{equation}

The equations~(\ref{pe_trans_final}), written as
$$
\left\{
\begin{array}{l}
\displaystyle \varepsilon'/c = \gamma(V)\, \big( \varepsilon/c - p_x\, V/c \big) \\
\ \\
\displaystyle p_x^\prime = \gamma(V)\, \big[ p_x - (\varepsilon/c) \, V/c \big] \qquad \quad p_y^\prime = p_y
\qquad \quad p_z^\prime = p_z
\end{array}
\right.
$$
and Lorentz transformations, such as
$$
\left\{
\begin{array}{l}
\displaystyle c t' = \gamma(V)\, \big( c t - x \, V/c \big) \\
\ \\
\displaystyle x^\prime = \gamma(V)\, \big[ x - (ct) \, V/c \big] \qquad \quad y^\prime = y
\qquad \quad z^\prime = z
\end{array}
\right.
$$
indicate that the quadruples $(ct, x, y, z)$ and $(\epsilon/c, p_x, p_y, p_z)$, known as Minkowski four-vectors, transform in the same way under changes of inertial reference frames.

\section{The laws of conservation of momentum and energy}

Terms such as additivity, conserved quantities, and conservation laws are not always understood in the same way. All these terms can refer to properties associated with individual particles or with a system of particles. In the present text, additivity does not simply mean that the quantity satisfies an additive composition rule. Instead, it means that if a property is associated with each particle in a system, the system as a whole can be described by an additive composition where each term corresponds to a single particle, and vice versa.

Thus, for a system of interacting particles, while momentum is an additive quantity -- meaning the system's momentum is the sum of the momenta of each particle -- energy does not satisfy additivity.
The system's energy results from the sum of the kinetic energies of each particle and the interaction potential energies between the particles. Although it results from an additive composition, the terms corresponding to interaction potential energies are associated with more than one particle. Energy is additive only when the particles hardly interact,
which occurs when the distances between them are such that interactions can be neglected. Only in such cases is the system's energy equal to the sum of the individual energies of the particles.

For example, when a proton approaches another proton at rest on a collision course with a kinetic energy on the order of $10^6$~eV,\footnote{\, 1~eV = $1.6 \times 10^{-19}$~J (joule).} the electromagnetic interaction potential energy -- even at distances on the order of atomic dimensions ($10^{-10}$~m) -- is on the order of 10~eV; this is indeed negligible compared to the energies of the incident proton and the proton at rest. Only at distances on the order of nuclear dimensions ($10^{-15}$~m) is the interaction energy between protons comparable to the energy of the incident proton.\footnote{\, The so called residual strong nuclear interaction between protons, although more intense, is short-range; it manifests only when the energy of the incident proton is sufficient to allow for separations smaller than nuclear dimensions.}

\begin{figure}[htbp]
\centerline{\includegraphics[width=7cm]{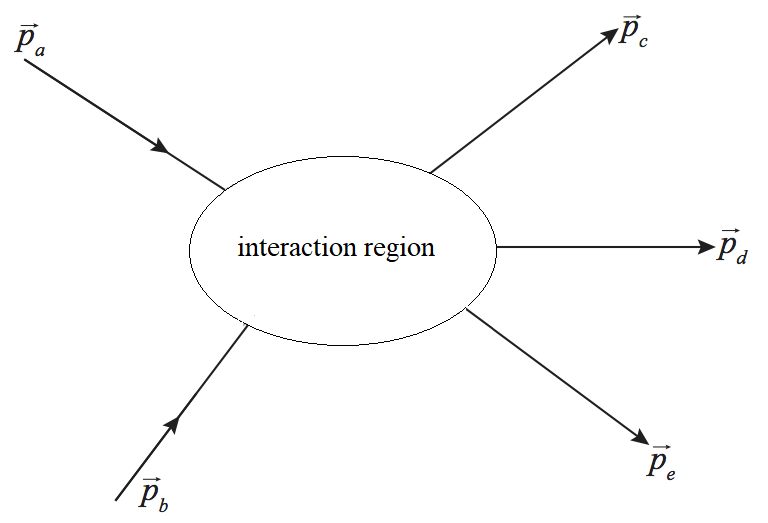}}
\caption{\small High-energy collision of two particles.}
\label{col_energ}
\end{figure}

Particle ``collision'' is, in reality, a process occurring over a very brief time interval within a very small \textit{interaction region}  (Figure~\ref{col_energ}). In this region, low-energy particles ($pc \ll mc^2$) may simply scatter, whereas high-energy particles ($\varepsilon \gg mc^2$) may undergo annihilation, leading to the creation of other particles.\footnote{\, Even at low energies, the final particles in a collision may differ from the initial ones, as they can coalesce or fragment.} Under these circumstances, interactions occurring before and after the collision can be neglected, and both the system's momentum and energy are governed by the so-called \textit{conservation laws}.

A quantity associated with a system of particles is said to be conserved, or to obey a conservation law, when it remains unchanged over time,\footnote{\, Although constant in a given reference frame, properties such as magnitude or direction are not necessarily the same in another reference frame.} even though the contributions associated with the individual particles may vary. In the case of a collision, the conservation laws for momentum and energy refer to the fact that the resultant momentum and total energy associated with the particles involved in the process retain the same magnitudes -- and direction (in the case of momentum) at the beginning and at the end of the process.

While in classical mechanics the conservation of energy and momentum for isolated systems derives from Newton's laws,  special relativity requires these conservation laws to be compatible with its fundamental principles.

Strictly speaking, for collision processes, it suffices to adopt the law of conservation of momentum. Just as the conservation of mass in low-energy collisions can be deduced from Galilean transformations by imposing momentum conservation, the conservation of energy follows naturally by imposing Lorentz invariance on the law of conservation of momentum.

\index{collisions! high-energy particle}
Consider the collision of two particles ($a, b$) that initially have momenta $\big( \vec p_a,\, \vec p_b \big)$ and energies $\big( \epsilon_a,\, \epsilon_b \big)$ in a reference frame $S$ (Figure~\ref{col_energ}).

If, for frame $S$ as well, the interaction in a given region of space results in three particles ($c, d, e$) with momenta and energies given, in that order, by $\big( \vec p_c,\, \vec p_d,\, \vec p_e \big)$ and $\big( \epsilon_c,\, \epsilon_d ,\, \epsilon_e \big)$, the conservation of momentum is expressed as
\begin{equation} \label{p_conserv}
\vec p_a + \vec p_b  =  \vec p_c + \vec p_d + \vec p_e
\end{equation}

If momentum conservation also holds in any other reference frame $S^\prime$ -- moving in uniform rectilinear translation with velocity $\vec V$ relative to $S$ --
\begin{equation} \label{pl_conserv}
\vec p_a^{\ \prime} + \vec p_b^{\ \prime}  =  \vec p_c^{\ \prime} + \vec p_d^{\ \prime} + \vec p_e^{\ \prime}
\end{equation}
then the additivity of momentum and energy implies, according to the transformation equations (\ref{pe_trans_final}), that for the momentum components parallel to $\vec V$,\footnote{\, In the non-relativistic regime, since the particle velocities ($\vec v$ and $\vec v^{\ \prime}$) in the two reference frames ($S$ and $S'$) satisfy the relation $\vec v^{\ \prime} = \vec v  - \vec V$, momentum additivity implies
$$\left\{
\begin{array}{l}
\displaystyle \big(  p_{a}^{\ \prime} +  p_{b}^{\ \prime} \big)_{\parallel} = \big(  p_{a} +  p_{b} \big)_{\parallel} - \big(  m_{a} +  m_{b} \big) V \\
\ \\
\displaystyle \big(  p_{c}^{\ \prime} +  p_{d}^{\ \prime} + \ p_{e}^{\ \prime} \big)_{\parallel} = \big(  p_{c} +  p_{d} +  p_{e} \big)_{\parallel} - \big( m_c + m_d + m_e \big)  V
\end{array}
\right.
$$
\indent In this case, the conservation of momentum is valid in both reference frames only if mass is additive and, likewise, obeys Lavoisier's law of conservation.
$$   m_{a} +  m_{b} = m_c + m_d + m_e $$
}
$$
\underbrace{\big(  p_{a} +  p_{b} \big)_{\parallel} - \big( \epsilon_a + \epsilon_b \big)  V/c^2}_{(p_a^\prime + p_b^\prime)_\parallel}
 = \underbrace{\big(  p_{c} +  p_{d} +  p_{e} \big)_{\parallel} - \big( \epsilon_c + \epsilon_d + \epsilon_e \big)  V/c^2}_{(p_c^\prime + p_d^\prime + p_e^\prime)_\parallel}
$$

Since, according to equation~(\ref{p_conserv}),
$$ \big(  p_{a} +  p_{b} \big)_{\parallel} =
\big( p_{c} + p_{d} +  p_{e} \big)_{\parallel }
$$
we obtain the law of conservation of energy in $S$,
\begin{equation} \label{conserv_S}
\epsilon_a + \epsilon_b  =
\epsilon_c + \epsilon_d + \epsilon_e
\end{equation}

On the other hand, expressing equation~(\ref{conserv_S}) in terms of the measurements made by an observer in $S^\prime$,
$$  \big( \epsilon_a^{\prime} + \epsilon_b^{\prime} \big) +\big(  p_{a}^{\ \prime} +
p_{b}^{\ \prime} \big)_{\parallel}  V  =
\big( \epsilon_c^{\prime} + \epsilon_d^{\prime} +  \epsilon_e^{\prime} \big) +
\big(  p_{c}^{\ \prime} +  p_{d}^{\ \prime} +  p_{e}^{\ \prime} \big)_{\parallel} V
$$
and taking into account that, according to equation~(\ref{pl_conserv}),
$$ \big(  p_{a}^{\ \prime} +  p_{b}^{\ \prime} \big)_{\parallel} =
\big(  p_{c}^{\ \prime} +  p_{d}^{\ \prime} +
\ p_{e}^{\ \prime} \big)_{\parallel}
$$
one obtains the energy conservation law in $S^\prime$,
\begin{equation} \label{conserv_Sl}
\epsilon_a^{\prime} + \epsilon_b^{\prime}  =
\epsilon_c^{\prime} + \epsilon_d^{\prime} + \epsilon_e^{\prime}
\end{equation}

The laws of conservation of energy and momentum transcend the limits of Classical Mechanics and Electromagnetism. Their validity extends to all phenomena in Particle Physics, even those involving particle creation and annihilation at high energies.

A significant portion of our knowledge regarding subatomic particles stems from the analysis of experiments involving the collision or scattering of particle beams, particularly of electrons and protons, in accelerators, where the particles are  generated with controllable energy and intensity, and directed at target nuclei or against other beams.

\section{The non-additivity of mass}

Perhaps the most famous equation in physics is the one expressing what Einstein termed the rest energy of a particle of mass $m$,

\vspace*{-0.3cm}
\begin{center}
\begin{tcolorbox}[width=2.4cm, colback=gray!15, colframe=gray!60,  arc=3mm, boxrule=0.5pt, top=1mm, bottom=1mm, left=3mm, right=3mm]
$\varepsilon_\circ = m c^2 $
\end{tcolorbox}
\end{center}


\vspace*{-0.1cm}
The content of this equation has been popularized as a mass-energy equivalence principle.
In this context, statements such as ``mass and energy are equivalent'',  ``matter can convert into energy'',  and ``energy can transform into matter''   have been propagated. Since then, the idea that Einstein's equation implies the possibility of the ``conversion of matter (or mass) into energy'' -- and vice versa -- has been propagated in the vast majority of publications on Special Relativity (with very rare exceptions~\cite{Okun, Nivaldo, Oguri, Morin}) and does nothing to clarify the changes Einstein actually introduced to the concepts of mass and energy.

Mass and energy do not have an independent physical existence; rather, they are attributes that can be associated with a given physical system. Therefore, energy cannot be transformed into matter, nor vice versa. While energy can be associated with any particle -- whether material or non-material, such as the electron and the photon -- mass is an attribute of material (massive) particles. The photon possesses no mass. According to Einstein's expression, mass and energy are correlated yet distinct attributes. The fact that they are correlated does not mean they are equivalent.

Given that mass is not strictly conserved in all processes, while energy follows a strict conservation law, what is the true meaning of converting mass into energy?

In an electron-positron annihilation process -- resulting in the creation of a pair of photons from a head-on collision where each material particle initially possesses the same energy\footnote{\, Since the electron and the positron have the same mass and, in this case, the same energy, the magnitude of their momenta is also equal.} -- the energy of each photon equals the energy of the initial electron or positron. Thus, it is correct to state that a zero-mass system was created from a material system while conserving energy, but incorrect to say that matter was entirely converted into energy.

Considering processes such as nuclear fission, in which a nucleus decays into a lighter nucleus and an alpha particle, the kinetic energy ($\varepsilon_c$) of the alpha particle is given by
$$ \varepsilon_c = \Delta m\, c^2 $$
where $\Delta m$ is the difference between the mass of the initial nucleus and the sum of the masses of the final nucleus and the $\alpha$ particle.

In this case, figuratively speaking, one can say that part of the rest energy associated with the mass of the initial nucleus was converted into the kinetic energy of the $\alpha$ particle.


Misinterpretations of the relationship between mass and energy originated with Einstein himself, starting with his 1905 paper [3], in which he states:

\vspace*{0.3cm}
\centerline{ ``\textit{If a body emits energy ($\varepsilon$) in the form of radiation, its mass decreases by $\varepsilon/c^2$.}''}
\vspace*{0.2cm}




Since energy is merely an attribute -- strictly speaking, a body does not emit energy -- Einstein's statement can be understood as a metaphor that is clarified by applying the principles of conservation of energy and momentum to the process he himself devised.

\vspace*{0.5cm}
\begin{center}
\begin{tcolorbox}[width=12cm, colback=gray!15, colframe=gray!60, title= \centerline{\large Process devised by Einstein}, coltitle=white!60,  arc=5mm, boxrule=0.5pt, top=3mm, bottom=3mm, left=4mm, right=4mm]
\hspace*{0.2cm} A body of mass $m$ emits two electromagnetic pulses in opposite directions, each with energy $\varepsilon/2$, relative to a reference frame in which the body is at rest. According to the principle of conservation of energy,
$$ (\mbox{\small initial energy}) \ \  m c^2 =  \varepsilon^\prime + \varepsilon \ \ (\mbox{\small final energy}) $$
where $\varepsilon^\prime$ is the energy of the body after the emission.

\vspace*{0.3cm}
\hspace*{0.2cm} Since the initial momentum is zero, and the momenta of the pulses have the same magnitude and opposite directions, the principle of conservation of momentum dictates that the final momentum of the body is also zero; thus, the body remains at rest with energy $\varepsilon^\prime = m^\prime c^2$.
This yields
$$ m c^2 = m^\prime c^2 + \varepsilon \qquad \Longrightarrow \qquad \big(m - m^\prime \big) = \frac{\varepsilon}{c^2}$$
indicating that the body's mass decreases by the amount
$$ \Delta m =  \varepsilon/c^2$$

\hspace*{0.2cm} In other words, one can say that part of the rest energy of a material body has been transferred to a non-material system (a pair of photons), and that this energy is related to the body's loss of mass.
\end{tcolorbox}
\end{center}

\vspace*{0.3cm}
The hypothesis adopted by Einstein -- that the expression for rest energy is valid for extended bodies as well -- not only reinforces the idea that mass depends on a body's internal energy (the kinetic and potential energy of its constituents) but also reveals another aspect: its non-additivity.

Consider that the rest energy ($E_\circ$) of a non-elementary particle of mass $M$ -- such as an atom, an atomic nucleus, a neutron, or a proton -- is given by the expression
$$ E_\circ = M c^2$$
and, furthermore, that in a reference frame where the particle is at rest, this rest energy arises from the energies ($\gamma_i m_i c^2$) of its constituents (with masses $m_i$) and the potential energy ($U_{\text{int}}$) due to the interactions between them,\footnote{\, $\gamma_i = \big(1- v_i^2/c^2\big)^{-1/2}$, where $v_i$ are the velocities of the constituent particles in the reference frame where the particle is at rest.}
$$ E_\circ = M c^2 = \sum_i \gamma_i m_i c^2 + U_{\text{int}}$$

Thus, it can be established that the mass of a system is not equal to the sum of the individual masses of its constituents. In other words, since it depends on a body's internal energy, mass is not an additive quantity.
\begin{center}
\begin{tcolorbox}[width=6.cm, colback=gray!15, colframe=gray!60, arc=3mm, boxrule=0.5pt, top=-1mm, bottom=1mm, left=3mm, right=3mm]
$$ M \ (\mbox{\small body}) \, \neq \, \sum_i m_i  \ (\mbox{\small constituents}) $$
\end{tcolorbox}
\end{center}



\subsection{Atomic and nuclear systems}

In general, for atomic and, especially, in nuclear phenomena, since the kinetic energies ($T_i = m_i v_i^2/2$) of the particles are much smaller than the rest energies ($\varepsilon_{{_0}i}=m_i c^2 $), a semi-relativistic limit is often used for the Lorentz factor:
$$ \gamma (v) = \lim_{v/c \ll 1} \Big( 1 - v^2/c^2 \Big)^{-1/2}
\simeq 1 + \frac{1}{2}
\frac{v^2}{c^2} \qquad \qquad (\mbox{semi-relativistic limit}) $$
such that the rest energy is written as
\begin{equation}\label{e-repouso}
E_\circ = M c^2 = \Big(\sum_i m_i \Big) c^2 \, + \, T_{\mbox{\tiny int}}
\, + \, U_{\mbox{\tiny int}}
\end{equation}
where $\displaystyle T_{\mbox{\tiny int}} = \sum_i \frac{m_i v_i^2}{2}$ is the total kinetic energy of the system's particles, or the internal kinetic energy of the system.

\begin{figure}[htbp]
\vspace{-0.2cm}
\centerline{\includegraphics[width=7.8cm]{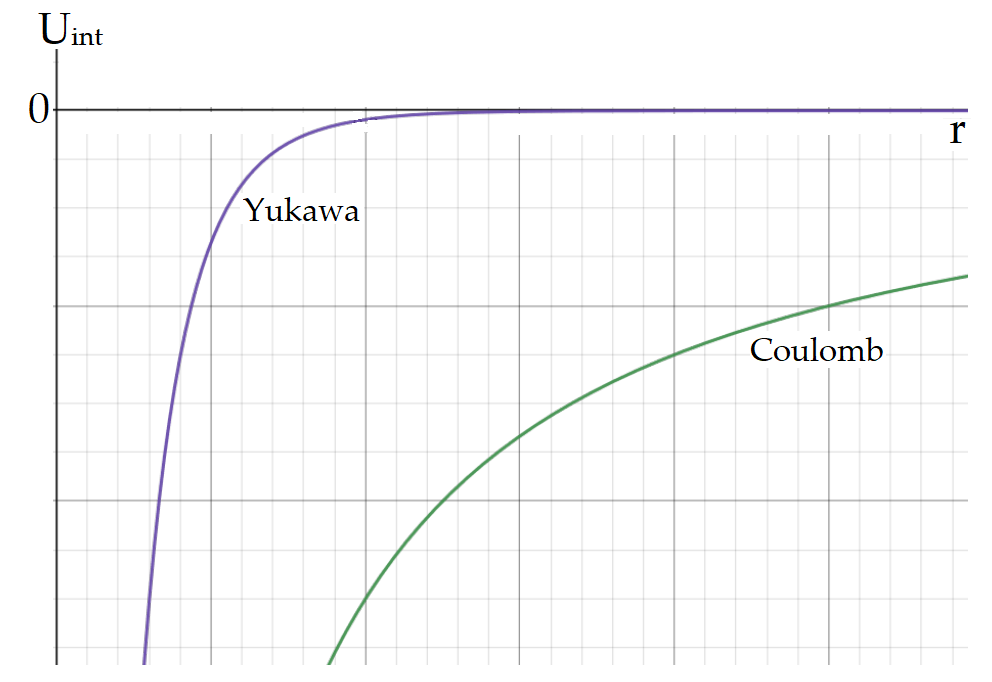}}
\vspace*{-0.3cm}
\caption{\small Coulomb and Yukawa interaction potential energies.}
\label{Yukawa_Coulomb}
\end{figure}

Classically, the interaction potential energy depends only on the instantaneous configuration of a system's constituents; relativistically, due to the finite propagation speed of any disturbance, interactions also depend on the particles' previous configurations. Therefore, strictly speaking, these interactions cannot even be expressed in terms of a potential energy. However, as a first approximation, the interactions between particles are expressed by a negative potential energy -- Coulomb-like ($- 1/r$) for atoms and Yukawa-like ($- e^{- r}/r$) for nuclei, where $r$ is the distance between the particles. Thus, the forces between the system's constituent particles are attractive (Figure~\ref{Yukawa_Coulomb}). These approximations reflect the fact that the interaction is long-range in atoms and short-range in nuclei.

On the other hand, according to Quantum Mechanics, equation~(\ref{e-repouso}) cannot be valid for atomic and nuclear systems. In such domains, the constituent particles of these systems do not possess definite momentum and energy. According to the Heisenberg's uncertainty principle, the smaller the confinement volume of the particles, the greater the uncertainties associated with momentum and energy. However, in principle, one may consider that the terms in equation~(\ref{e-repouso}) represent the average values of the particles' kinetic and potential energies.

In these systems, especially in nuclei, the magnitude of the potential energy exceeds that of the kinetic energy. Therefore, for $T_{\mbox{\tiny int}} < |U_{\mbox{\tiny int}}|$, and in accordance with equation~(\ref{e-repouso}), the system's mass ($M$) is less than the sum of the masses $\displaystyle \Big( \sum_i m_i \Big)$ of its constituents; that is,
\vspace*{-0.3cm}
\begin{center}
\begin{tcolorbox}[width=6.cm, colback=gray!15, colframe=gray!60,  arc=3mm, boxrule=0.5pt, top=-1mm, bottom=1mm, left=3mm, right=3mm]
$$ M  \, < \, \sum_i m_i  \quad (\mbox{\small atoms and nuclei}) $$
\end{tcolorbox}
\end{center}


\vspace*{-0.2cm}
The difference $\displaystyle \Delta M = \sum_i m_i - M$, termed the \textit{mass defect}, determines the system's \textit{binding energy} ($\varepsilon_\ell$),
\vspace*{-0.2cm}
$$\varepsilon_\ell = \Delta M \, c^2$$
which represents the minimum amount of energy that must be transferred to a system to decompose it into its constituents.

While for atomic or molecular systems -- bound by electromagnetic interactions -- the binding energy is on the order of $10$~eV and the ratio of mass defect to system mass ($\Delta M/M$) is less than $10^{-8}$ (practically negligible), for nuclear systems -- bound by residual strong interactions -- the ratio is on the order of $10^{-3}$ to $10^{-2}$, given that the interaction potential energy is also negative, and the nuclear binding energy is much higher, on the order of $1$~MeV (Table~1).

\renewcommand{\arraystretch}{1.5}
\begin{table}[htbp]
\caption{Comparison of mass defect for different nuclei  }
\vspace{0.2cm}
\centering{\begin{tabular}{lccc}
\hline
nucleus & mass defect ($\Delta M$) & mass ($M$) & ratio ($\Delta M/M$)  \\
\hline
deuteron ($^2\mbox{\tt H}$)   & $0.002388~\mbox{\tt u}$ &  \quad $2.013553~\mbox{\tt u}$   & $1.18 \times 10^{-3}$
\\
iron ($^{56}\mbox{\tt Fe}$)   & $0.528460~\mbox{\tt u}$ & \ \ \! $55.920680~\mbox{\tt u}$  & $9.45 \times 10^{-3}$ \\
uranium ($^{235}\mbox{\tt U}$) & $1.915020~\mbox{\tt u}$ & $234.993467~\mbox{\tt u}$ & $8.15 \times 10^{-3}$ \\
\hline
\end{tabular}} \\
\centering{$1\,\mbox{\tt u} = 931.5$~MeV/$c^2$ (atomic mass unit)}
\label{defeito_massa}
\end{table}
\renewcommand{\arraystretch}{1}

It is therefore possible to state:
\begin{center}
\begin{tcolorbox}[width=13.5cm, colback=gray!15, colframe=gray!60,  arc=5mm, boxrule=0.5pt, top=2mm, bottom=2mm, left=4mm, right=4mm]
The mass of a nucleus is less than the sum of the masses of its protons and neutrons.
\end{tcolorbox}
\end{center}


If we consider Einstein's expression for rest energy to be equally valid for particles with internal structure -- such as protons, neutrons, nuclei, atoms, and molecules -- then, in addition to the issue of non-additivity, the dependence of mass on the kinetic and potential interaction energies of their constituents also becomes apparent. Thus, one must understand the correlation between mass and energy established by Einstein. Therefore, it can be said that a portion of the mass of atoms and molecules is of electromagnetic origin, corroborating—at least in part—M. Abraham's intuition from 1902.


\subsection{Subnuclear systems}

Subnuclear systems are defined as non-elementary particles -- such as the \textit{proton} and the \textit{neutron}, which belong to a broad family known as \textit{baryons}, and particles like the \textit{pion} and the \textit{kaon}, which belong to another family known as \textit{mesons}. These particles, both baryons and mesons, collectively termed \textit{hadrons}, may carry a positive, negative, or zero electric charge and are composed of so-called \textit{quarks}, which are bound together by one of nature's fundamental interactions: the \textit{strong interaction}. However, baryons are composed of three quarks, and mesons are composed of a quark-antiquark pair.

Quarks -- particles considered elementary -- constitute one of the two fundamental families of the Standard Model of Particle Physics and possess electric charges that are fractions ($\pm 2/3, \pm 1/3$) of the elementary charge ($e$).\footnote{\, The quark family is composed of 6 species or flavors: up ($u$), down ($d$), strange ($s$), charm ($c$), bottom ($b$), and top ($t$), and their respective antiquarks; $\bar u$, $\bar d$, $\bar s$, $\bar c$, $\bar b$, and $\bar t$; while the other family of elementary particles -- the so-called leptons -- includes the electron ($e$), the muon ($\mu$), the tau ($\tau$), their three respective neutrinos ($\nu$), and their six respective antiparticles.}  Particle Physics, or High Energy Physics, is a field predominantly conducted in laboratories housing large particle accelerators\footnote{\, Such as the LHC (Large Hadron Collider) at CERN (European Laboratory for Particle Physics) in Geneva, Switzerland.} that investigate phenomena occurring during particle collisions at energies far exceeding the particles' rest energies.
In this field, relativistic kinematics is employed -- encompassing the laws of conservation of energy and momentum in their full scope -- and, to date, it has not challenged any of the consequences of the principles of Special Relativity.


\begin{figure}[htbp]
\centerline{\includegraphics[width=8.8cm]{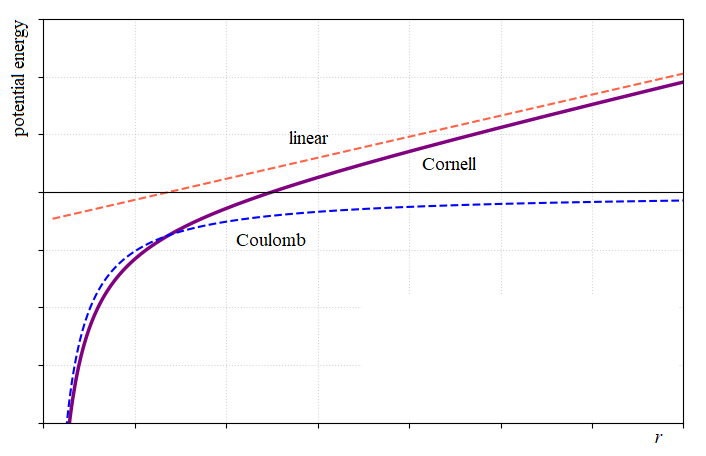}}
\vspace*{-0.2cm}
\caption{\small Cornell potential energy of quark interaction.}
\label{quark_potencial}
\end{figure}

As a first approximation, the strong interaction between quarks can be described by the so-called Cornell potential, which is Coulomb-type when the distance between the quarks is almost zero ($\sim 0,\!1$~fm), growing linearly and positively with distance from a given value (Figure ~\ref{quark_potencial}). This interaction energy of the type $ - K /r \, + \, \sigma r \, + c $,\footnote{\, $K$, $\sigma$ and $c$ are experimentally determined parameters (appendix ~\ref{ap}).} has the character of an infinite potential well, being compatible with the hypothesis that at very short distances quarks behave as free particles (asymptotic freedom), but confined within hadrons.

According to Heisenberg's principle, confined within a region of linear dimensions ($a$) on the order of $10^{-15}$~m, the minimum uncertainty in the momentum of a quark is on the order of
\[\Delta p\simeq \frac{\hbar}{a} \sim 200\text{\ MeV}/c\]
where $\hbar \simeq 1.055 \times 10^{-34} \text{\ J.s}$ is the reduced Planck constant.

Thus, both the kinetic term ($pc$) and the interaction potential are much larger than the rest energies of the lightest quarks, $u$ and $d$ ($m_u = 2.2 MeV/c^2$ and $m_d = 4.7 MeV/c^2$)~\cite{takahashi}.
Therefore, unlike what occurs in atoms and nuclei, the masses of the proton and neutron, collectively called nucleons, are much greater than those of their constituent particles, the $u$ and $d$ quarks.\footnote{\,  While the proton ($p$) is composed of two $u$ quarks and one $d$ quark ($uud$), the neutron ($n$) is composed of one $u$ quark and two $d$ quarks ($udd$).}

\begin{center}
\begin{tcolorbox}[width=12.5cm, colback=gray!15, colframe=gray!60,  arc=5mm, boxrule=0.5pt, top=2mm, bottom=2mm, left=4mm, right=4mm]
The mass of nucleons is much greater than the sum of the masses of its quarks.
\end{tcolorbox}
\end{center}



By analogy with electromagnetic interactions -- where the action of the fields responsible for interactions between electrically charged particles is associated with photons (massless particles acting as mediators of electromagnetic interactions) -- the theory of strong interactions in particle physics, Quantum Chromodynamics (QCD), characterizes interactions between quarks as being mediated by similarly massless particles: gluons.

Since the mass of a nucleon is on the order of $10^3$~MeV/$c^2$ while the sum of the masses of the three quarks that compose them is about 10~MeV/$c^2$, in principle these constituents would contribute no more than 1\% of the total mass of these hadrons. More precise calculations~\cite{Yang}, which consider the existence of virtual quark-antiquark pairs heavier than the $u$ and $d$ quarks -- due to quantum fluctuations, even if ephemeral --, indicate that the mass of quarks, instead of 1\%, contributes about 9\% to the mass of the proton, while 91\% comes from the kinetic energy of the quarks (31\%) and the energy fluctuations associated with the mediators of the strong interactions, the gluons (60\%).

In the case of baryons and mesons heavier than nucleons, the difference between the mass of the hadron and the masses of its constituents is not as pronounced (Table \ref{hadron_comparison}).


\renewcommand{\arraystretch}{1.2}
\begin{table}[htbp]
\centering
\caption{Comparison of the masses of different hadrons~\cite{takahashi}}
\label{hadron_comparison}
\vspace{0.2cm}
\begin{tabular}{lcccc}
\hline
 \textit{hadron} & quark & hadron & sum of quark  & mass fraction \\
             & content & mass (Gev/$c^2$)  & masses (Gev/$c^2$) & from quarks  \\
\hline
\textit{meson}  & & & & \\
neutral pion  ($\pi^0$) & $u\bar{u}$  & $0.135$ & $ 0.004$ & $\ \, 3.0\%$ \\
kaon ($K^+$) & $u\bar{s}$ & $0.494$ & $ 0.095$ & $19.2\%$ \\
J/Psi ($J/\psi$) & $c\bar{c}$ & $3.097$ & $ 2.546$ & $82.2\%$ \\
\hline
\textit{barion} & & & & \\
charmed lambda ($\Lambda_c^+$) & $udc$ & $2.286$ & $ 1.280$ & $56.0\%$ \\
omega  ($\Omega_{ccc}^{++}$) & $ccc$ & $4.793$ & $ 3.819$ & $79.7\%$ \\
\hline
\end{tabular}
\end{table}
\renewcommand{\arraystretch}{1.0}

\begin{center}
\begin{tcolorbox}[width=6cm, colback=gray!15, colframe=gray!60,  arc=3mm, boxrule=0.5pt, top=-1mm, bottom=1mm, left=3mm, right=3mm]
$$ M \ (\mbox{\small hadron}) \, > \, \sum_i m_i \  \ (\mbox{\small \textit{quarks}})$$
\end{tcolorbox}
\end{center}

Despite these characteristics of hadron masses, considering that hydrogen constitutes the predominant fraction of baryonic matter in the "known" Universe (approximately 93\%), and that the mass of the proton does not result from the mass of its ponderable constituents -- the $u$ and $d$ quarks --, but rather from the energy associated with the dynamics of these quarks, as a consequence of Einstein's hypothesis, it can be inferred -- in light of Special Relativity -- that the mass of the observable cosmos is not directly determined by matter.

\appendix
\section{Estimates}\label{ap}

\vspace*{-0.2cm}
Typical experimentally adjusted values for the parameters of the Cornell potential are
\[K = 0.102 \quad \text{e} \quad \sigma = 0.91~\text{ GeV/fm}    \quad c= -0.25\]

\vspace*{-0.2cm}
Thus, the potential is given by
\[ V(r)=-\frac{0.102}{r}+ 0.91\ \, r - 0.25\qquad \quad V (\text{GeV}), \quad r (\text{fm}) \]
and the magnitude of the force exerted between the two quarks, given by the gradient of the potential,
\[  F(r)= \frac{{\rm d} V}{{\rm d}r}   = \frac{0.102 }{r^{2}}+ 0.91  \quad (\text{GeV/fm}) \]

\vspace*{-0.2cm}
If the quarks are separated by more than 1~fm, the strong force becomes practically constant and extremely intense,
\[ F \simeq 1\text{\ GeV/fm} \simeq 1.6 \times 10^5\text{\ newtons} \]

\vspace*{-0.2cm}
\noindent while the electromagnetic interaction force is only about \( 10^2\text{\ newtons} \).


\end{document}